\documentclass[letterpaper, 10 pt, conference]{ieeeconf} 
\IEEEoverridecommandlockouts
\usepackage{cite}
\usepackage{amsmath,amssymb,amsfonts}
\usepackage{algorithmic}
\usepackage{graphicx}
\usepackage{textcomp}
\usepackage{xcolor}
\def\BibTeX{{\rm B\kern-.05em{\sc i\kern-.025em b}\kern-.08em
    T\kern-.1667em\lower.7ex\hbox{E}\kern-.125emX}}
\begin{document}

\title{EEG-Based Auditory BCI for Communication in a Completely Locked-In Patient Using Volitional Frequency Band Modulation\\

}

\author{Deland Liu$^{1}$, Frigyes Samuel Racz$^{2}$, Zoe Lalji$^{1}$ and José del R. Millán$^{1,2,3}$%
\thanks{$^{1}$Chandra Family Department of Electrical and Computer
Engineering, $^{2}$Department of Neurology, $^{3}$Department of Biomedical Engineering, The University of Texas at Austin, Austin, TX, USA. E-mail:
        {\tt\small deland.liu@utexas.edu}, {\tt\small  jose.millan@austin.utexas.edu}}
}

\maketitle

\begin{abstract}
Patients with amyotrophic lateral sclerosis (ALS) in the completely locked-in state (CLIS) can lose all reliable motor control and are left without any means of communication. It remains unknown whether non-invasive electroencephalogram (EEG)-based brain–computer interfaces (BCIs) can support volitional communication in CLIS. Here, we show that a CLIS patient was able to operate an EEG-based BCI across multiple online sessions to respond to both general knowledge and personally relevant assistive questions. The patient delivered “Yes”/“No” responses by volitionally modulating alpha and beta band power at different channels, guided by real-time auditory feedback from the BCI. The patient communicated assistive needs above chance in all sessions, achieving a perfect score in the final session. Performance on general knowledge questions varied across sessions, with two sessions showing accurate and above-chance responses, while the first and last sessions remained at chance level. The patient also showed consistent modulation patterns over time. These findings suggest that non-invasive BCIs may offer a potential pathway for restoring basic communication in CLIS.
\end{abstract}


\section*{Disclosure}
© 2025 IEEE. Personal use of this material is permitted. Permission from IEEE must be obtained for all other uses, in any current or future media, including reprinting/republishing this material for advertising or promotional purposes, creating new collective works, for resale or redistribution to servers or lists, or reuse of any copyrighted component of this work in other works.

\section{Introduction}
Amyotrophic lateral sclerosis (ALS) is a progressive and devastating neurodegenerative disease that primarily affects motor neurons, leading to the gradual loss of voluntary muscle control, including respiratory and bulbar functions \cite{masrori2020amyotrophic}. 
In later stages, many patients lose all voluntary muscle movements, including the ability to move their eyes or their fingers—the final remaining means of interaction with their environments—resulting in a condition known as the completely locked-in state (CLIS) \cite{chaudhary2016brain}. For patients in this state, all conventional forms of communication, such as eye tracker, switches and head-mounted laser become inaccessible, presenting profound challenges to autonomy and quality of life.

In such cases, brain–computer interfaces (BCIs) offer a promising alternative by translating neural activity into control signals for direct communication with the external environment.
 Non-invasive electroencephalogram (EEG)-based BCIs, such as P300 speller systems, have been demonstrated in ALS patients with preserved visual function, achieving communication rates up to 11.20 bits per minute \cite{mccane2015p300}. 
 Speier et al. \cite{speier2017online} augmented this system with a language model, reaching 46.07 bits per minute in locked-in patients with reliable gaze control. 
Locked-in ALS patients have also used implanted, intracortical or electrocorticography (ECoG)-based BCI systems. 
Milekovic et al. \cite{milekovic2018stable} demonstrated home-based use of a local field potential-based BCI over 138 days, achieving 6.88 correct characters per minute in a patient with some mobility left.  
Pels et al. \cite{pels2019stability} reported consistently high performance, achieving 91\% correct hits in a target-selection task using an ECoG-based BCI, maintained over a three-year period in a patient with intact visual function.
Another ECoG-based BCI enabled a patient with residual eye movements left to communicate at a rate of two letters per minute \cite{vansteensel2016fully}. 

The development of BCIs for ALS patients has predominantly focused on patients in the locked-in state (LIS), where some residual muscle control (e.g., eye or mouth movements) remains and cognitive functions might be largely intact. 
In contrast, there is limited evidence that implanted or non-invasive BCI systems can enable reliable communication in patients with CLIS \cite{mcfarland2020brain, marchetti2015brain}. Several factors have been proposed to explain this challenge in cases of complete paralysis, including visual impairments as well as the inability to maintain gaze fixation \cite{marchetti2015brain}, the potential absence of goal-directed cognition \cite{kubler2008brain}, and altered evoked potentials\cite{chaudhary2022spelling}, which are commonly used as BCI control signals.
To date, only one published study has demonstrated BCI-based communication in a CLIS patient, which used two implanted 64-channel microelectrode arrays in the supplementary and primary motor cortices \cite{chaudhary2022spelling}. The BCI system provided real-time auditory feedback by mapping neural firing rates to tone frequency, which the patient modulated volitionally. “Yes”/“No” responses were conveyed by maintaining the tone frequency within predefined high or low thresholds, enabling the patient to select letters and words at an average rate of approximately 5 bits per minute.
While this implanted approach shows promise, it remains unknown whether non-invasive EEG-based BCIs can support volitional control and communication in CLIS patients.

In this study, we demonstrate that a patient in CLIS, with no voluntary muscle movement, no reliable eye control, and an inability to sustain eye opening or gaze fixation, was able to volitionally control an EEG-based BCI to communicate. Prior to using the BCI, the patient had no reliable means of communication. Using the BCI system, however, the patient reliably communicated assistive needs and answered general knowledge questions through binary (“Yes”/ “No”) responses, with variable performance across online sessions. Our system enabled control through patient-driven modulation of alpha and beta power over specific channels. Increases in these band powers signaled a “Yes” response, while baseline band powers indicated a “No”.
To accommodate visual impairment, the BCI provided continuous auditory feedback, with tone frequency indicating the predicted class and volume reflecting classifier confidence. These findings provide rare evidence that non-invasive EEG-based BCIs can enable reliable communication in CLIS.

\section{Methods}
\subsection{Patient}
The patient (male, 58 years old, CLIS) had an official diagnosis of ALS in November 2017. He lost speech in August 2018 and mobility in October 2018. He was on mechanical ventilation since July 2020, received nutrition through a percutaneous endoscopic gastrostomy tube since May 2018, and stayed at home with caregiver support since his diagnosis in 2017 until he passed away in September 2024. 
Patient used the Tobii Dynavox eye-gaze device from May 2018 until October 2020 when his ocular function began to decline, for text-to-speech communication as well as internet and social media access. After that, the patient tried several different versions of EMG and switches-based text-to-speech communication, which were unsuccessful. Following this, the patient's primary means of communication involved a letter board and binary responses conveyed through movements of his eyebrows, eyes, or cheek muscles. This mode was relatively successful for a few years. However, it became increasingly sporadic and inconsistent as his disease progressed, and there was no form of muscle control left until we started BCI experiments with the patient in March 2024. 
There was no clinical diagnosis that the patient was completely locked-in.
Patient's family member provided written informed consent for the experiment, in accordance with the Declaration of Helsinki. The experimental protocol was approved by the local ethics committee (2020-03-0073, The University of Texas at Austin, TX, USA). 

\subsection{Experimental paradigm and set-up}
A total of seven sessions were conducted with the patient over several weeks, each lasting approximately 45 minutes to maintain engagement and minimize fatigue. These included three offline sessions and four online sessions. The first two offline sessions were conducted on consecutive days, while the third took place three weeks later. The first online session was conducted three days after the final offline session. The remaining three online sessions were spaced approximately 5, 10, and 14 days apart, respectively.

Each offline session consisted of 3-4 runs, where the patient was auditorily and pseudo-randomly cued to either volitionally increase EEG amplitude in the alpha and beta bands (“Modulated” class) or return it to baseline (“Baseline” class).
Each offline run contained 14-20 trials, evenly split between classes across sessions. Data from all offline sessions were used to train the subject-specific binary classifier.

Each online session consisted of a combination of three possible run types: up to one offline run, one or two standard online runs, and up to one online assistive run (see Section ``Assistive Task'' below).
 In standard and assistive online runs, the patient used the BCI to answer “Yes”/“No” questions by modulating EEG activity—“Modulated” signaled “Yes,” and “Baseline” signaled “No.” Real-time auditory feedback reflected ongoing BCI control.
 
Standard online runs had 6–10 trials. The second online session used a balanced 50/50 class split of “Modulated”/“Yes” vs “Baseline”/“No”, whereas the first, third, and fourth sessions followed a 60/40 split. The slight class imbalance resulted from early termination of some runs or sessions in response to signs of patient disengagement (e.g., yawning). This also led to the variation in the number of runs and trials across sessions.
Online assistive runs were included in the second through fourth sessions—containing 5, 4, and 4 trials respectively. The class distributions were 60/40, 50/50, and 75/25, respectively. Class imbalance was not controlled in these runs, as the patient’s intended responses to predefined assistive questions were unknown beforehand and were provided retrospectively by the caretaker at the end of each trial.
Offline runs recorded during online sessions served as warm-up periods to reacquaint the patient with the modulation task. 

Throughout the recording, scalp EEG was recorded at 512 Hz using an eego system (ANT Neuro, Netherlands) from 32 electrodes located at FP1, FPz, FP2, F7, F3, Fz, F4, F8, FC5, FC1, FC2, FC6, M1, T7, C3, Cz, C4, T8, M2, CP5, CP1, CP2, CP6, P7, P3, Pz, P4, P8, POz, O1, Oz, O2 in 10/20 international coordinates. The ground electrode was placed at AFz and the reference electrode was placed at CPz.

\subsection{Brain-computer interface task design} 
Although all task cues/instructions were delivered auditorily, real-time EEG signals from selected channels were visually displayed during offline sessions for the patient. These signals were band-pass filtered between 7–30 Hz to encompass the alpha and beta frequency bands. The display was intended to provide the patient with visual feedback on the effectiveness of his mental modulation strategy. 

During offline sessions, the patient was seated in front of the screen (14-inch display, 2560 $\times$ 1440 pixels, 60~Hz refresh rate, ThinkPad X1 Carbon), which was carefully positioned by the caretaker to ensure proper visibility. Sessions were scheduled during periods when the patient could comfortably sustain eye opening, with durations kept short to accommodate their limitations.

Each offline trial began with a 3 s inter-trial interval, followed by an 8 s rest period, accompanied by a brief 0.5 s audio cue saying “Rest”. The trial number was then played (approx. 1 s), followed 2 s later by an auditory instruction cue: either “Make it noisy” or “Make it clean” (approx. 1 s). These corresponded to the “Modulated”/“Yes” and “Baseline”/“No” classes, respectively. Colloquial terms were intentionally used to ensure the patient could understand the task. 
After the trial cue, a tone corresponding to the trial class was played: 200 Hz for “Baseline” trials and 370 Hz for “Modulated” trials. This tone gradually increased in volume over a 10 s period to simulate continuous auditory BCI feedback; both tones were matched in overall volume. Looking at the feedback from real-time EEG signals, the patient was instructed to perform the mental strategy corresponding to the cue, aiming to either increase or decrease the absolute amplitude of the EEG signal displayed on the screen as long as the auditory feedback was played.
 The instructions were “When instructed “Make it noisy”, please try to assume a mental state/find a strategy to increase the absolute amplitudes of the EEG signals” and “When instructed “Make it clean”, please try to assume a mental state/find a strategy to decrease the absolute amplitudes of the EEG signals to baseline or flat”.

In standard online runs, each trial began with a 3 s inter-trial interval followed by an 8 s rest period. The question number was then played (approx. 1 s), and 2 s later, a general knowledge “Yes”/“No” question was presented auditorily (approx. 1–4 s). 3 s after the question, a bell tone (~0.5 s) was played to cue the patient to begin BCI control by responding either “Yes” (corresponding to the “Modulated” class) or “No” (corresponding to the “Baseline” class), using the same mental strategy to increase or decrease EEG amplitudes in the alpha and beta bands as practiced during the offline sessions.
Auditory feedback began 4 s after the bell cue. As in the offline sessions, the tones representing each class were the same (200 Hz for “Baseline”/“No” class and 370 Hz for “Modulated”/“Yes” class). However, in the online setting, the volume and tone were continuously adjusted based on the accumulated posterior probability of the class detected by the BCI. As a result, the tone that increased in volume over time reflected the class currently favored by the system. 

Two decision thresholds were defined for the accumulated evidence. If one threshold was crossed, an additional auditory feedback message (“You selected Yes” or “You selected No”, approx. 1 s) was played to indicate the decoded command. If neither threshold was reached within 20 s, a time-out occurred and a short “Time out” audio cue (approx. 0.5 s) was presented, marking the end of the trial.

The general knowledge “Yes”/“No” questions included basic facts spanning domains such as natural science, biology, technology, and food (e.g., Do human hearts beat? Does a year have 1000 days?). There was no visual display of EEG signals in online sessions. 

\subsection{Assistive task}
To evaluate the practical utility of the BCI, online assistive runs were conducted in which the patient responded to personalized “Yes”/“No” questions provided by the caretaker. In the second and third online sessions, assistive questions were organized into a decision tree structure, where each follow-up question depended on the preceding question and patient’s response to the preceding one.
Each run contained 4 to 5 questions, corresponding to 4 or 5 layers in the decision tree. For example, if the patient responded “No” to the question “Are you comfortable?”, a possible follow-up would be “Do you have pain in your back?”.
This tree-based structure allowed the patient to communicate a rich and specific message by the end of the run. 
In contrast, assistive questions in the first session were unstructured and not part of a sequential decision tree.

After each question, the caretaker rated their confidence in the patient's response on a subjective 5-point scale, with 5 indicating high confidence and 1 indicating low confidence. These ratings were based on the caretaker’s familiarity with the patient and contextual factors such as time of day or behavioral cues observed during the session. Confidence scores greater than 3 were considered correct responses; scores of 3 or below were considered incorrect.

\subsection{“Modulated” vs “Baseline” classifier}
EEG signals were first bandpass filtered between 7–30 Hz using a causal, second-order Butterworth filter. Alpha (7–12 Hz) and beta (12–30 Hz) band power features were extracted from all channels using 2 s sliding windows updated every 1 s. Power was computed by squaring the filtered signals and normalized per trial using the average power from the final 1 s of the pre-trial rest period.

Alpha and beta band features were concatenated across channels to form the full feature set. A Random Forest-based feature selection procedure was used to identify the most discriminative features. Specifically, a TreeBagger model with 100 trees was trained across 10 iterations within each run of a leave-one-run-out cross-validation framework. Feature importance scores were computed using out-of-bag permutation error, and the top 20 features were identified based on aggregated importance rankings across runs and iterations.

To optimize feature count, linear SVM models were trained using subsets of the top 20 features (ranging from 20 to 4), and the number yielding the highest average cross-validation accuracy was chosen. 
A final linear SVM model was then trained on the full offline dataset using the selected top-ranked features. To calibrate class probabilities, a logistic regression model was fitted to the SVM decision scores. The decoder remained fixed and was not updated during online sessions. 

\subsubsection{Evidence accumulation}
To issue a command (“Yes” or “No”), the BCI continuously integrates class probabilities over time. A decision is made when the accumulated probability for one class exceeds a set threshold. This accumulation process is performed using exponential smoothing \cite{leeb2013transferring}: 
\begin{equation}
    Prob_{i}= 0.95 \times Prob_{i-1} + 0.05 \times p_{i}
    \label{eq:evidence accumulation}
\end{equation}
where $Prob_{i}$ refers to the accumulated probability for a class up to the $i$-th sample, and $p_{i}$ indicates the posterior probability of the class at $i$-th sample. 

\subsection{BCI performance analysis}
To evaluate online BCI performance, we considered the accuracy of the BCI, level of control, command delivery performance, and latency to command delivery. 
To evaluate the BCI's accuracy, we used the Cohen's Kappa value (Kappa) \cite{cohen1968weighted}, which is suitable for evaluating imbalanced data distributions, which can arise in online sessions. 
\begin{equation}
        Kappa=\frac{p_{a} - p_{e}}{1 - p_{e}}
        \label{eq:Cohen}
\end{equation}
where $p_{a}$ is the classification accuracy and $p_{e}$ is the expected accuracy by chance.

To quantify the patient's level of control over the BCI, we computed bar dynamics, defined as the percentage of time during which the accumulated probability output by the decoder favored the correct class by surpassing the 50\% chance threshold.

To evaluate command delivery performance, we computed the percentage of correct responses the patient would have achieved under a fixed decision threshold of $60\%$ for each class. This provides a standardized estimate of BCI effectiveness that accounts for subjectivity in threshold settings during online sessions, which were sometimes set ambitiously and may have underestimated the patient's true control capability. 

Lastly, to further examine the practicality of the BCI system, we computed the latency to correct hit, defined as the time elapsed from the onset of BCI decoding to the moment the accumulated evidence for the correct class crossed the correct decision threshold.

\subsection{Physiological analysis}
To examine the spatial distribution of alpha and beta band activity for each class, average, baseline-normalized alpha and beta band power features were computed using a non-causal, second-order Butterworth filter to eliminate phase distortion.
For each session and frequency band, class-wise differences (“Modulated”/“Yes” – “Baseline”/“No”) were computed and visualized topographically.

\subsection{Statistical analysis}

Chance-level Kappa values were estimated by performing 10,000 random permutations of the test labels. A Kappa value was computed for each permutation, and their mean was used as the chance-level.

To identify, for each session and frequency band, the EEG channels that showed statistically significant higher alpha and beta band power features in “Modulated”/“Yes” class compared to the “Baseline”/“No” class, we performed a non-parametric, two-tailed permutation test independently at each EEG channel. Each sample corresponded to the alpha or beta band power extracted from a sliding window, as used during decoder training (offline) and operation (online). 

To perform the permutation test, samples from both classes were pooled and their labels were randomly shuffled 1,000 times to generate a null distribution of mean power differences for each channel. The p-value was computed as the proportion of permutations in which the absolute difference exceeded or equaled the observed difference in the original data. Benjamini–Hochberg procedure was used to correct for multiple comparisons across EEG channels

\section{Results}
\section{Performance of ALS Patient's Online BCI Responses to General Knowledge “Yes”/“No” Questions}

\begin{figure}[ht]
\centering
\includegraphics[width=\linewidth, height=\linewidth]{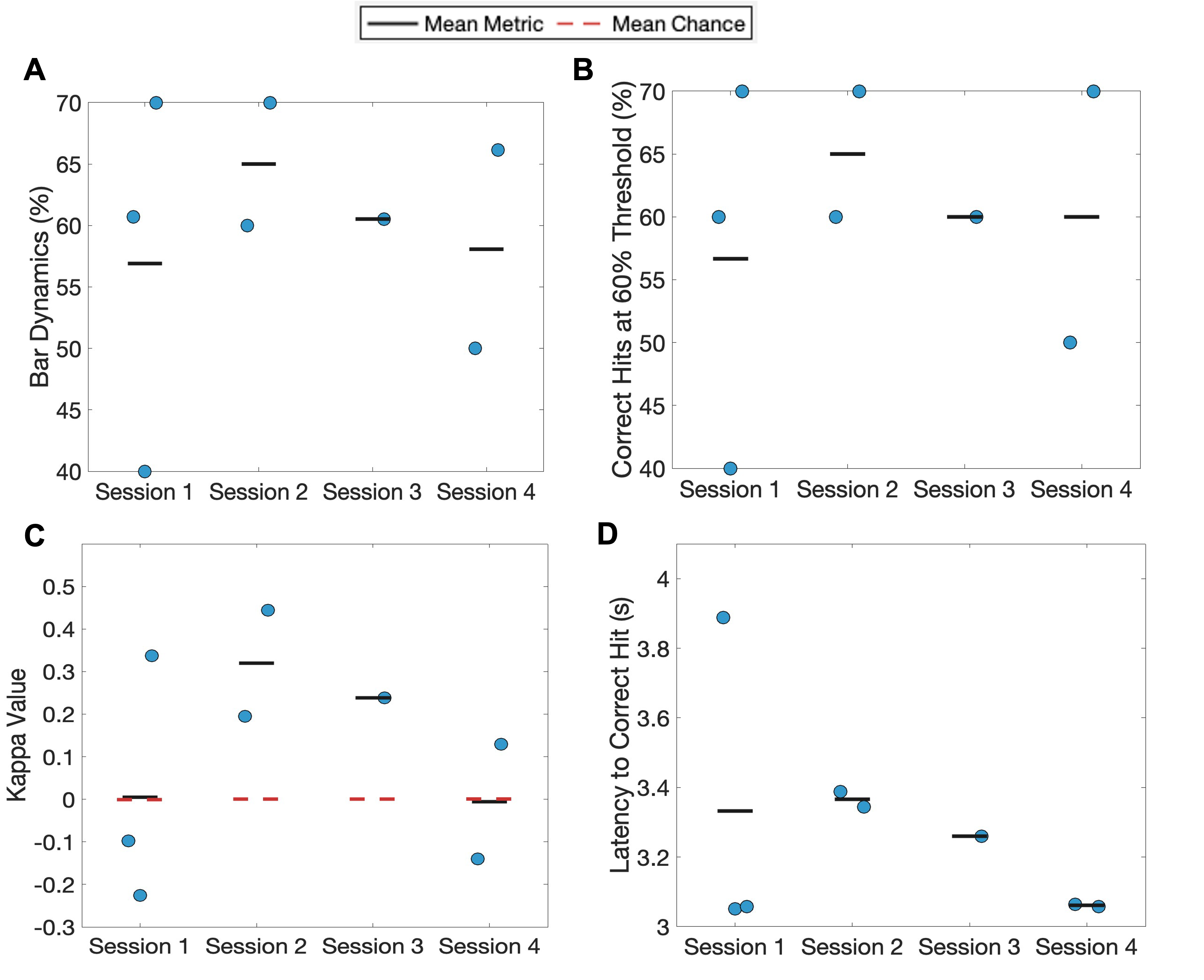}
\caption{\label{fig:Online_results}\textbf{Online BCI performance during responses to general knowledge “Yes”/“No” questions.} \textbf{A} Bar dynamics, \textbf{B} Percentage of correct hits at a 60\% threshold on the accumulated posterior probability, \textbf{C} Kappa value, and \textbf{D} Latency to a correct BCI hit across online sessions. Each dot represents the metric from a single run, and the solid black line indicates the mean across runs. Dashed red lines in \textbf{C} represent chance level performances. }
\end{figure}

To evaluate the patient's ability to respond to general knowledge questions via an online BCI, we assessed four metrics: Bar dynamics, Percentage of correct hits at a 60\% threshold on the accumulated posterior probability, Kappa value, and Latency to correct hits across the four online BCI sessions (Fig.\ref{fig:Online_results}). 
Despite some variability across runs, average bar dynamics remained above 55\% in all four sessions and exceeded 60\% in session 2 and session 3. In total, 6 out of 8 runs had the bar dynamics at or above 60\% (Fig.\ref{fig:Online_results}A). 
Kappa values were near chance level in sessions 1 and 4, but exceeded chance in sessions 2 and 3, reaching values above 0.20 (Fig.\ref{fig:Online_results}C).
Percentage of correct hits at the 60\% threshold began at 56.67\% in session 1 and remained at or above 60\% in all subsequent sessions. Across all runs, 6 out of 8 reached or exceeded the 60\% mark (Fig.\ref{fig:Online_results}B). The latency to a correct hit remained below 3.5 s across all runs (Fig.\ref{fig:Online_results}D). 

\section{Performance of ALS Patient's Online BCI Responses to Assistive “Yes”/“No” Questions}
\begin{figure}[ht]
\centering
\includegraphics[width=\linewidth, height=\linewidth]{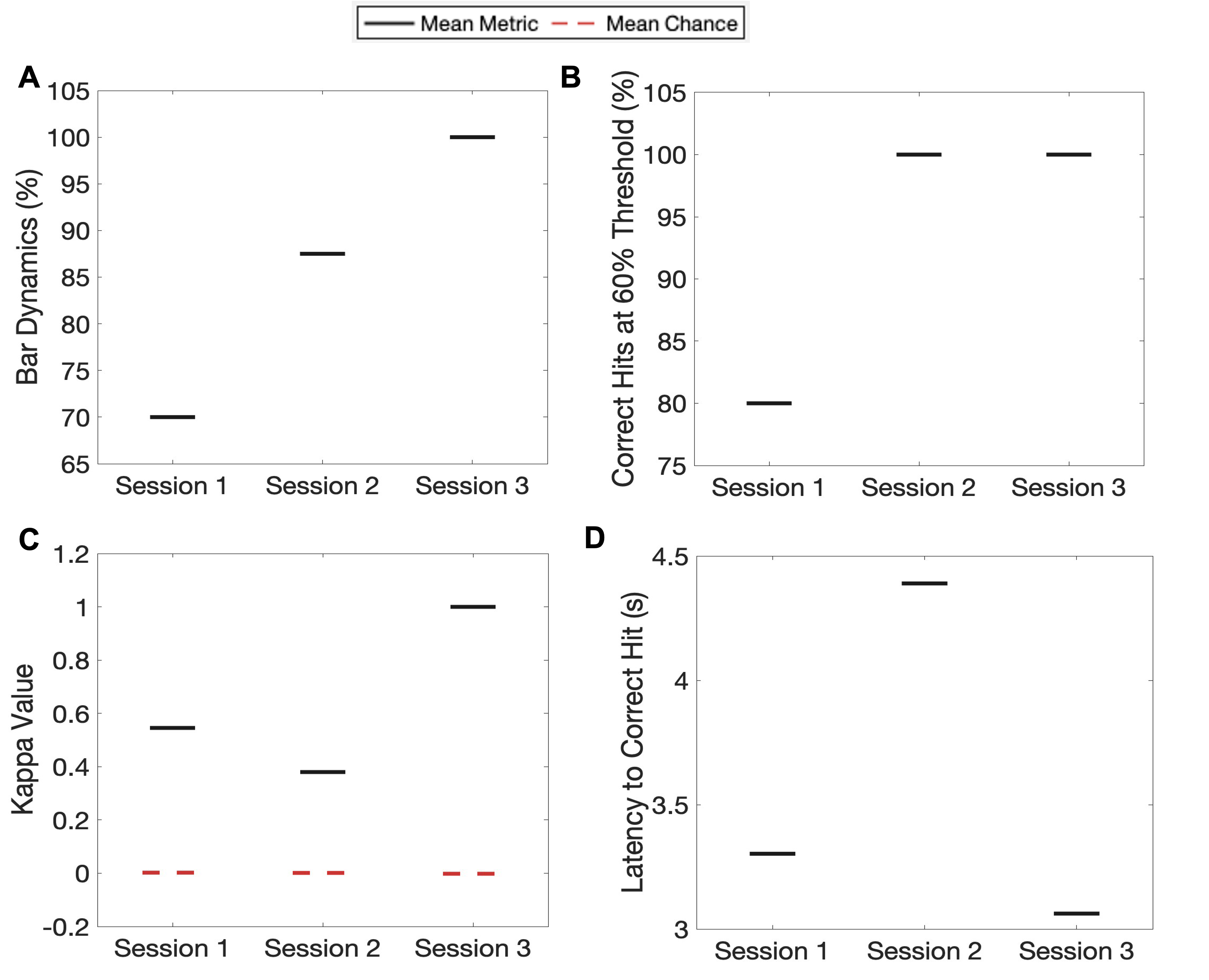}
\caption{\label{fig:Online_assistive_results}\textbf{Online BCI performance during responses to assistive “Yes”/“No” questions.} \textbf{A} Bar dynamics, \textbf{B} Percentage of correct hits at a 60\% threshold on the accumulated posterior probability, \textbf{C} Kappa value, and \textbf{D} Latency to a correct BCI hit across sessions.The solid black line indicates the metric value for the single run conducted in each session. Red dashed lines in \textbf{C} represent chance level performances. Note that each session consisted of a single assistive run. }
\end{figure}

We evaluated the same four performance metrics to assess the patient's ability to use the BCI system to respond to assistive questions.
Shown in Fig.\ref{fig:Online_assistive_results}A, bar dynamics remained consistently at or above 70\% across all three sessions, reaching 100\% in the final session. In the final session, all classification windows were correctly labeled, resulting in a perfect Kappa value (Fig.\ref{fig:Online_assistive_results}C). Kappa values in the first two sessions were also above chance, remaining at or above 0.40.
Percentage of correct hits at the 60\% threshold began at 80.0\% in the first session and reached 100\% in both the second and third sessions (Fig.\ref{fig:Online_assistive_results}B). The latency to correct hit remained below 4.5 s throughout (Fig.\ref{fig:Online_assistive_results}D).
By responding to the rule-based assistive questions, the patient communicated needs such as "I am not comfortable at the moment. I am experiencing pain in my back and would like a massage specifically at my lower back." in the last online session.  

\section{Volitional and Stable Modulation of Alpha and Beta Bands in the ALS Patient. }

\begin{figure}[ht]
\centering
\includegraphics[width=\linewidth]{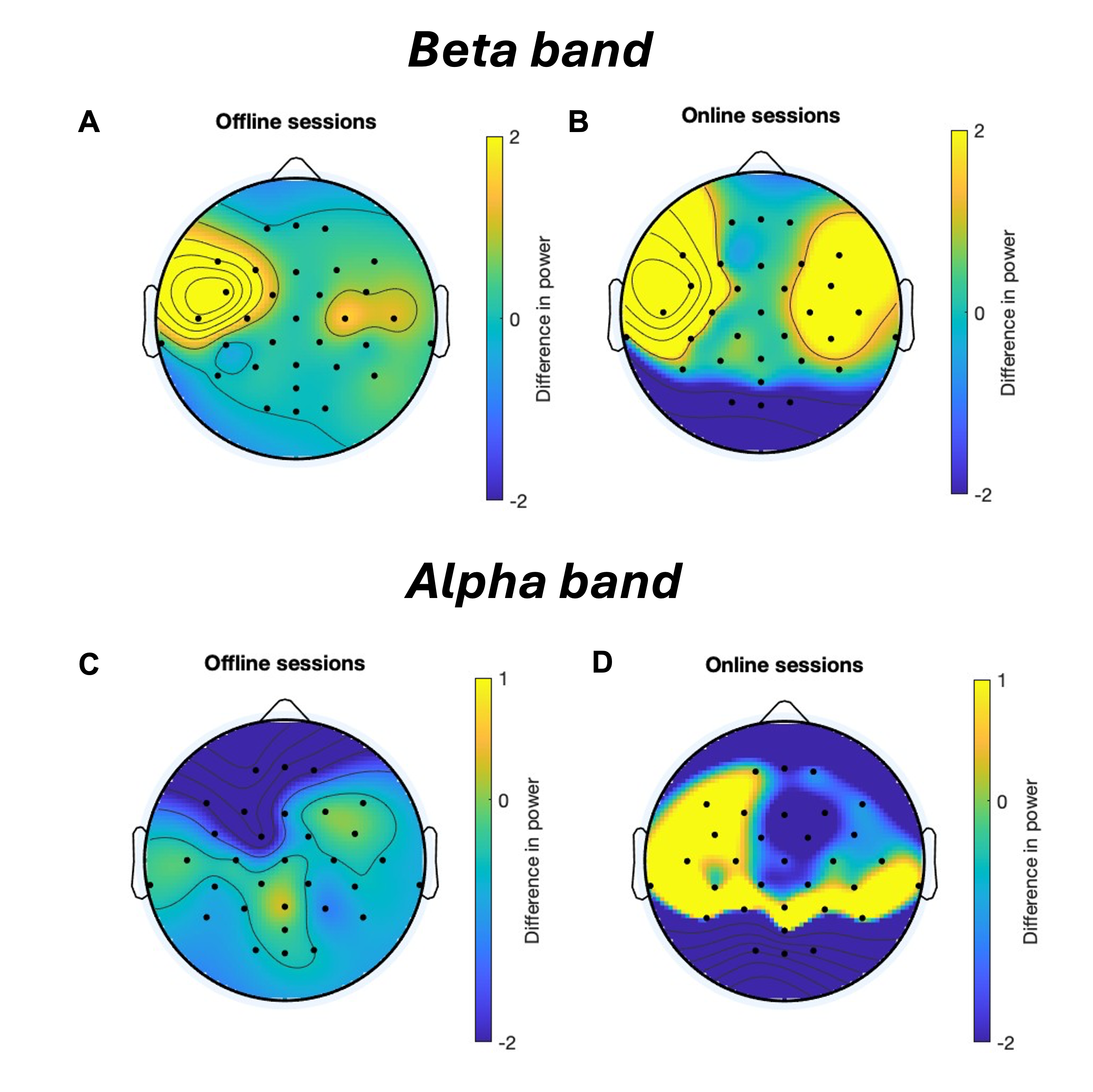}
\caption{\label{fig:topo}\textbf{Topographical differences in Alpha and Beta band power between the “Modulated” and “Baseline” classes.} \textbf{A} and \textbf{B} show beta band differences for offline and online sessions; \textbf{C} and \textbf{D} show the corresponding alpha band differences.}
\end{figure}

The spatial distribution of alpha and beta power differences between the “Modulated” and “Baseline” classes suggests that the patient was able to volitionally modulate neural activity in specific frequency bands and channels. During the offline session, beta power increased significantly in the “Modulated” class, compared to “Baseline” class, at multiple channels (Fig.\ref{fig:topo}A). A two-tailed permutation test with correction for multiple comparisons identified 13 channels with significant or near-significant increases in beta power, including F3, $p_{corrected}=0.0145$; F4, $p_{corrected}=0.0293$; F8, $p_{corrected}=0.1080$; FC5, $p_{corrected}=0.1360$; FC1, $p_{corrected}=0.0071$; C3, $p_{corrected}=0.0071$; Cz, $p_{corrected}=0.0046$; CP1, $p_{corrected}=0.0050$; CP2, $p_{corrected}=0.0050$ and P7, $p_{corrected}=0.0663$.
In the online sessions, this spatial pattern partially persisted. Several of the previously significant channels remained significant or trended toward significance (Fig.\ref{fig:topo}B), including F4, $p_{corrected}=0.0800$; F8, $p_{corrected}=0.0427$; FC5, $p_{corrected}=0.0427$; C3, $p_{corrected}=0.0427$; Cz, $p_{corrected}=0.0320$ and CP1, $p_{corrected}=0.0240$. The number of channels showing significant or near-significant increases in online sessions remained consistent with those observed during offline sessions.

In the alpha band, Fig.\ref{fig:topo}C shows increased modulation at Pz during the “Modulated” class in the offline session; however, this effect did not remain significant after correction for multiple comparisons ($p_{corrected}=0.1231$). A similar trend was observed in the online sessions (Fig.\ref{fig:topo}D), with Pz showing an increase ($p_{corrected}=0.1551$). Compared to the offline sessions, the online sessions revealed additional spatial engagement, with four more channels exhibiting near-significant increases in alpha power during the “Modulated” class—significant prior to corrections—including P4, $p_{corrected}=0.0996$; P8, $p_{corrected}=0.0960$; F7, $p_{corrected}=0.1500$ and T7, $p_{corrected}=0.0960$. 

\section{Discussion}
In this study, we demonstrated that a patient in CLIS was able to volitionally operate an EEG-based BCI across multiple online sessions by modulating activity in the alpha and beta frequency bands at specific channels. Using this system, the patient was able to communicate assistive needs with high accuracy, including achieving perfect control in the final session. The patient also used the BCI to respond to general knowledge questions, although performance varied —accuracy was at chance level in the first and last sessions, but exceeded chance and remained high in the two intermediate sessions.
In addition, the latencies to correct hits fall under 5 s, supporting the potential feasibility of the system for functional use. 
Lastly, no clear trend of improvement was observed across sessions in either the general knowledge or assistive communication paradigms.
Physiological analyses revealed consistent spatial patterns of significant or near-significant increases in beta-band power in the “Modulated” class compared to “Baseline” during offline and online sessions. In the alpha band, consistent, near-significant spatial patterns were also observed across sessions, with a broader distribution of alpha power increases in the “Modulated” condition during online sessions relative to offline.

The broader spatial engagement of alpha band activity observed during online sessions, compared to offline, may reflect a learned neural strategy facilitated by real-time BCI feedback. This interpretation aligns with the principles of operant conditioning, where feedback-driven learning induces neurophysiological changes \cite{perdikis2018cybathlon, liu2025brain}. 
Alpha power increases have been associated with enhanced inhibitory control and the suppression of task-irrelevant activity~\cite{mahjoory2019power, klimesch2007eeg, liu2025personalized}. In this context, given the auditory nature of the task, the patient may have volitionally increased alpha power in the parietal regions (e.g., P4 and P8) to suppress irrelevant visual processing in an effort to improve BCI control during online use. 

Interestingly, online BCI performance was higher during assistive runs than standard runs when patient responded to general knowledge questions. This may be due to the increased task engagement driven by the personal relevance of assistive questions, which may have facilitated greater volitional effort. Personal relevance has been shown to enhance attention, arousal and neural responses (e.g., increased event-related potentials) \cite{bayer2017impact}. 


The study has several limitations. Firstly, to minimize patient fatigue, each recording session had a limited number of runs and trials, especially in the online assistive runs. This limits the statistical power and generalizability of findings. In addition, due to time constraints in deploying the BCI for the patient, the BCI decoder was not fully optimized, leaving room for improvement in areas such as model selection and hyperparameter tuning to improve decoding performance.


In summary, this study demonstrates that a patient in CLIS, without any reliable means of communication, was able to volitionally modulate brain activity to answer questions and express needs using a non-invasive EEG-based BCI. These findings add to the limited body of evidence supporting BCI-based communication in CLIS. Further research is needed to assess whether similar control can be achieved across a broader population of CLIS individuals.

\bibliographystyle{IEEEtran}

\end{document}